\documentclass[twocolumn]{elsart}
\usepackage{natbib,graphicx}
\usepackage{amsmath,amsfonts,amssymb,color,graphicx}
\begin{document}
\runauthor{Maitra and Bailyn}
\begin{frontmatter}
\title{Low-Intensity States in the Neutron Star X-ray Binary Aql X-1}
\author[yale]{Dipankar Maitra},
\ead{maitra@astro.yale.edu}
\author[yale]{Charles Bailyn}
\address[yale]{Dept. of Astronomy, Yale University, New Haven, USA}

\begin{abstract}
Aql X-1 was observed in a Low Intensity State (LIS), a state that is usually 
characterized by high optical but low X-ray flux. Our daily monitoring of 
the source using the 1.3m telescope in Cerro Tololo Inter-America Observatory 
operated by the Small and Moderate Aperture Research Telescope System (SMARTS) 
was used to trigger a series of target-of-opportunity (ToO) observations of 
the source using the Rossi X-ray Timing Explorer satellite (RXTE). The X-ray 
colors and 
temporal variability studies suggest that the source was in a powerlaw 
dominated state featuring high rms variability in the lightcurve and dominance 
of hard photons 
during this LIS. The ToO observations were continued until the source 
made a transition to the canonical thermal dominated or high/soft state.
\end{abstract}

\end{frontmatter}

\begin{figure*}[ht]
\centering
\includegraphics[height=0.39\textwidth, width=0.7\textwidth]{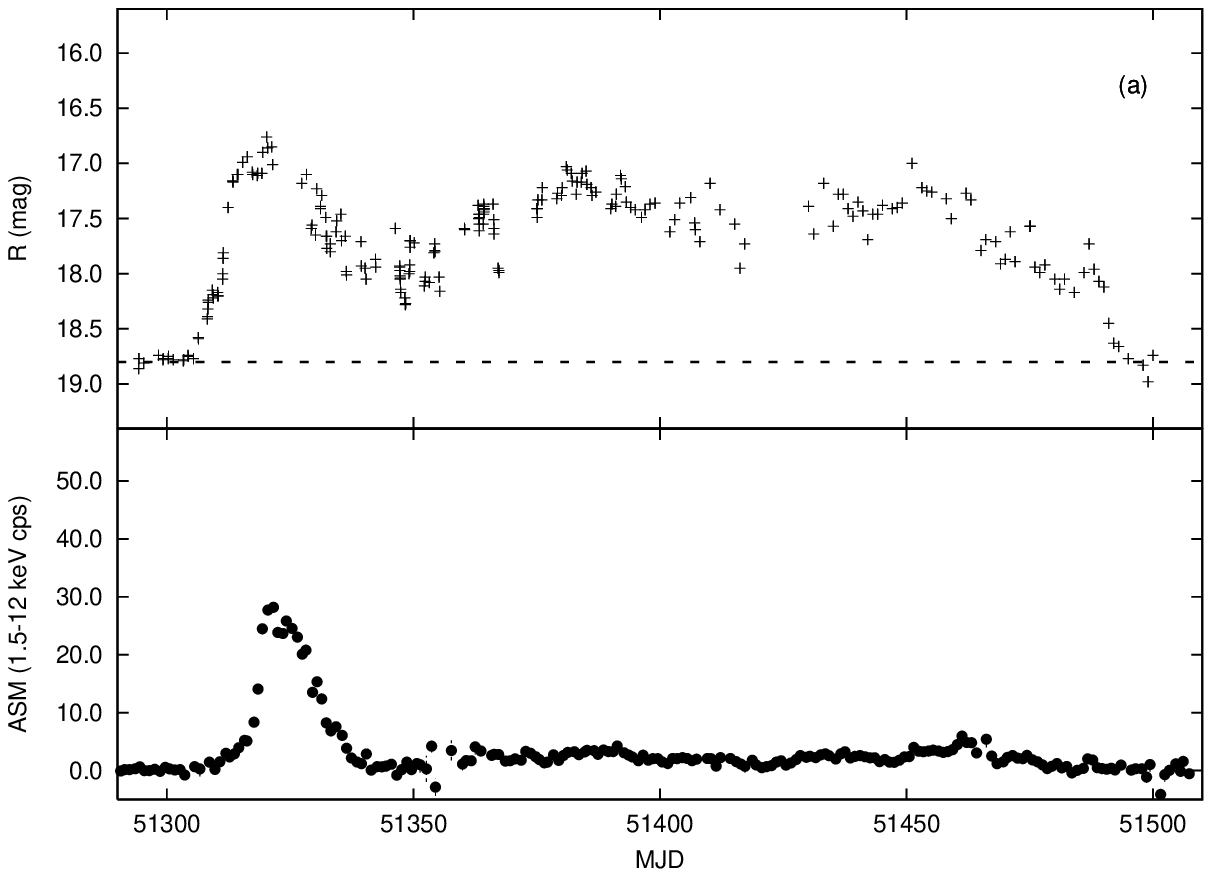}
\includegraphics[height=0.39\textwidth, width=0.7\textwidth]{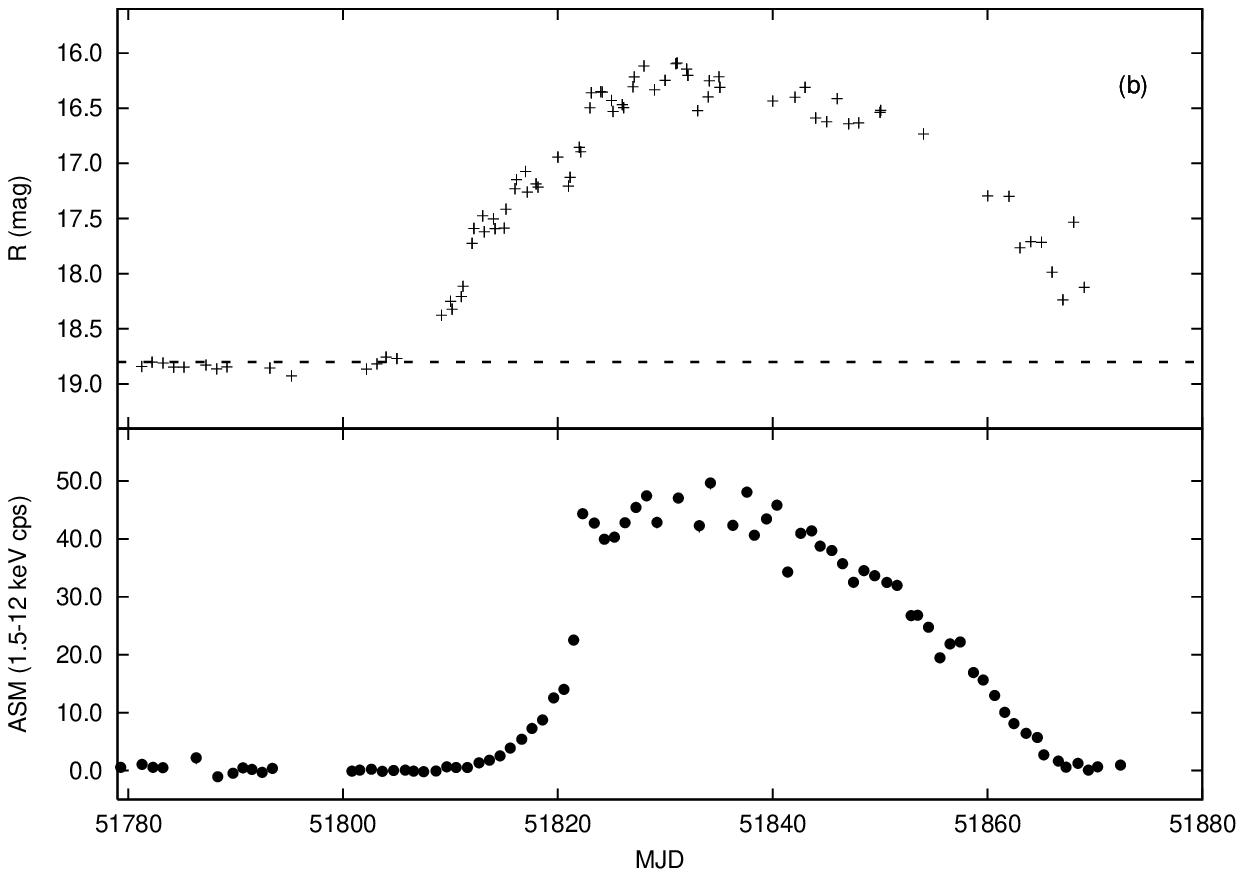}
\includegraphics[height=0.39\textwidth, width=0.7\textwidth]{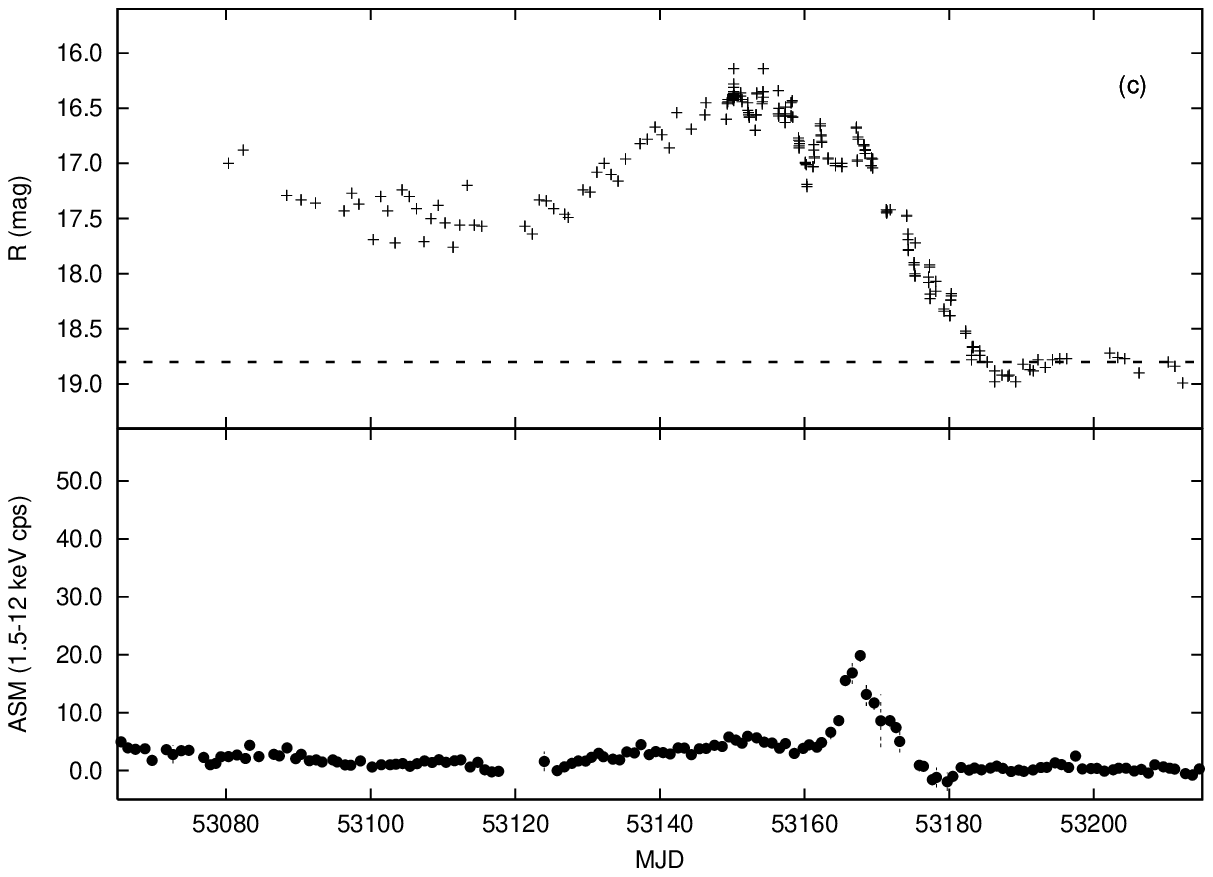}
\caption{
Different types of outburst in Aql X-1. In panels (a), (b) and (c) we show the 
R band as well as ASM activity during three outbursts. The R band data was 
taken using the SMARTS 1.3m telescope in CTIO, Chile. ASM data courtesy the 
ASM team at MIT.
Panel (a): A FRED followed by an extended LIS during 1998.
Panel (b): A normal outburst with FRED profile during 2000.
Panel (c): An extended LIS followed by a transition to the soft state during 
2004 (Maitra and Bailyn, in preparation). The lack of data during the end of 
(b) and the 
beginning of (c) is because the source was too close to the Sun to observe at 
night. The dotted line in the optical panel is the mean quiescent brightness of
Aql X-1. The quiescent X-ray flux is below ASM detection limits.
\label{lc}}
\end{figure*}

Aql X-1 is a well studied transient neutron star X-ray binary (NSXRB) system 
that goes into outburst about once every year. Since its launch, the 
RXTE has observed 5 major outbursts from 
this source that follow the typical Fast Rise and Exponential Decay (FRED) 
lightcurve morphology \cite{csl1997}. Besides these major outbursts, many small scale 
activities have been noted that were clearly non-FRED (see Fig.~\ref{lc}) . 
Regular monitoring in optical/IR (OIR) bands have revealed that the OIR flux 
during some of these minor X-ray activities can become comparable to the peak 
OIR flux during normal outbursts.

Detection of Aql X-1 in hard X-rays using the INTEGRAL satellite was reported 
by Molkov et al. \cite{atel259} and also in optical by Ilovaisky and Chevalier 
\cite{atel260}, after it came out from behind the Sun in mid-March 2004. During
mid-May 2004 it showed signs of coming out of this prolonged LIS \cite{atel279}
and entering a full outburst. Here we present the results of our pointed X-ray 
observations of the source during May-June 2004.

\begin{figure*}[ht]
\centering
\includegraphics[height=0.7\textwidth, width=1.0\textwidth]{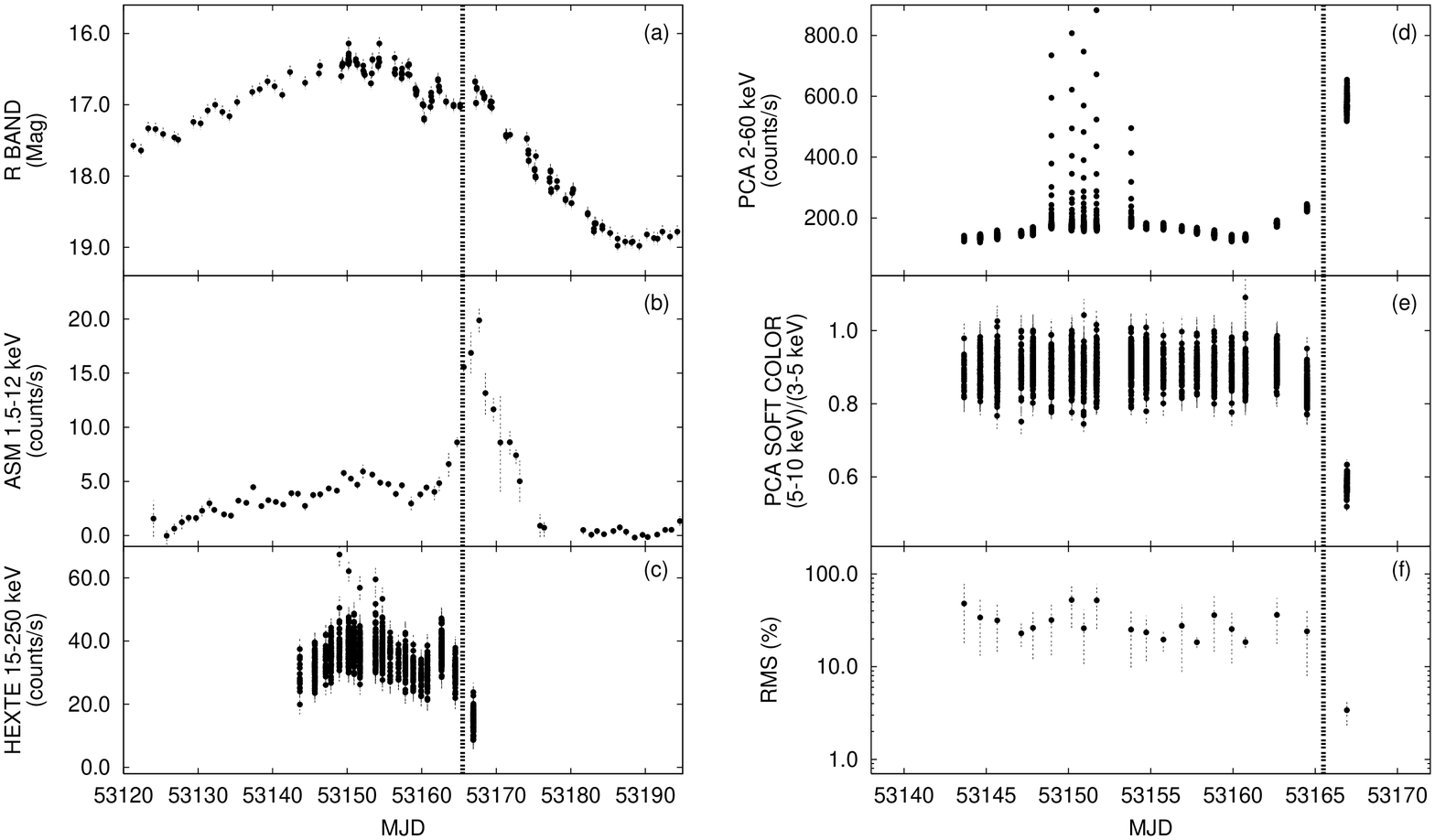}
\caption{
The evolution of lightcurve, colors and rms variability during the outburst of
June 2004. The dotted line marks the transition from a powerlaw dominated to a 
thermal dominated state.
Panel (a): Optical R band lightcurve from SMARTS.
Panel (b): RXTE ASM 1.5-12 keV X-ray lightcurve.
Panel (c): RXTE HEXTE ~15-250 keV hard X-ray lightcurve. Note that the HEXTE 
lightcurve peaks about the same time as the optical does. Also remarkable, is 
the sharp decline in HEXTE flux right after the transition to soft state.
Panel (d): RXTE PCA ~2-60 keV lightcurve. Note the frequent type I bursts from 
MJD 53147-53155 and the sharp rise in flux after the state transition.
Panel (e): The hardness of the source given by the ratio of counts in hard to 
soft band also drops abruptly, marking the transition from one state to 
another.
Panel (f): The rms variability of the incident photon flux between, in the frequency range of 0.01-10 Hz also shows sharp changes during state transition.
\label{hr}}
\end{figure*}

\begin{figure*}[ht]
\centering
\includegraphics[width=0.7\textwidth]{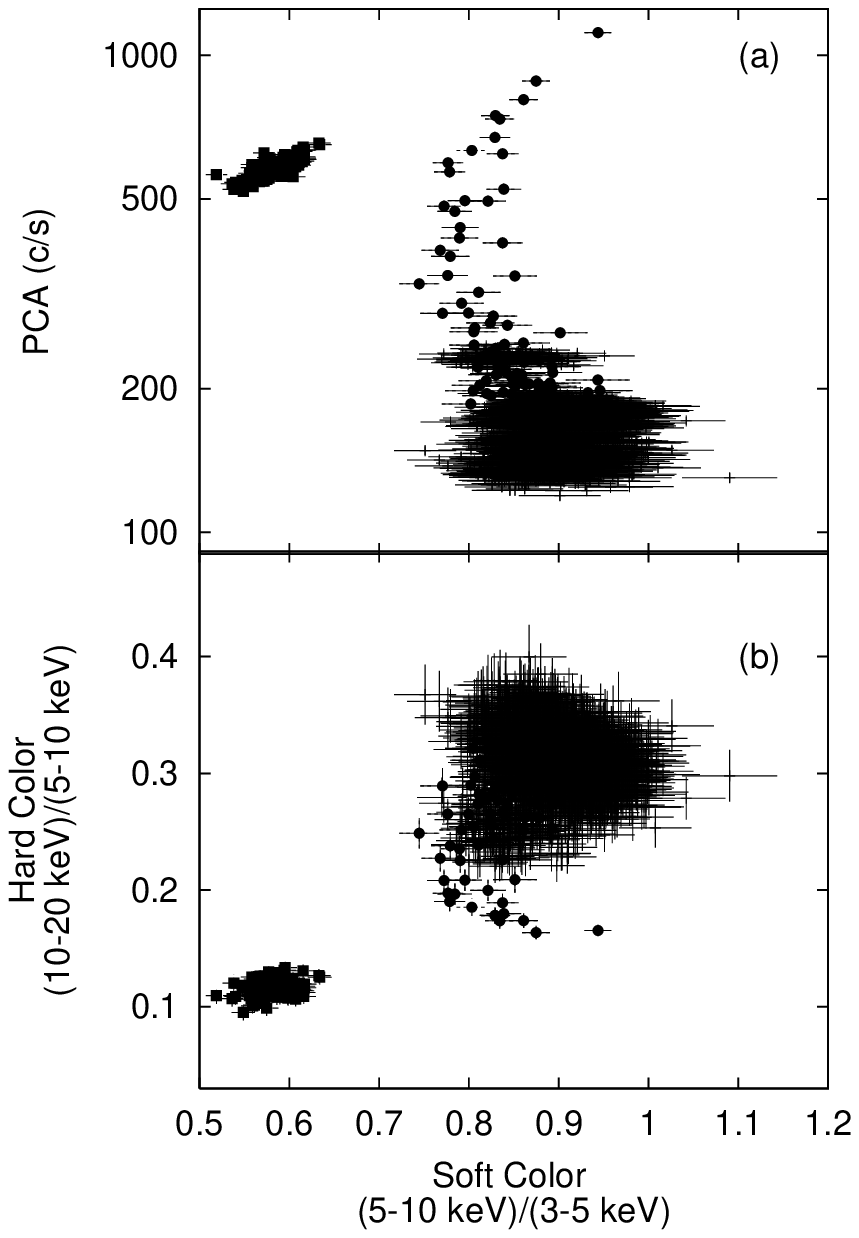}
\caption{
The color-luminosity and color-color diagrams respectively show clearly the two
distinct spectral nature of the source before and after the state transition. 
The soft state is the small island in the left side of both plots shown by 
filled squares ($\blacksquare$), the hard state, with more data-points and 
larger scatter occupy the island on the right side shown by plus symbols ($+$).
The filled black circles ($\bullet$) are data during type I bursts and hence 
are clear outliers compared to the rest of the data.
\label{cc}}
\end{figure*}

Preliminary study of the X-ray colors and variability 
(Fig.~\ref{hr} and Fig.~\ref{cc}) shows that during the LIS, the X-ray state is very similar to the canonical 
powerlaw dominated or low/hard state with high rms variability and frequent 
type I bursts. The total $\sim$ 2-60 keV flux remained at a fairly 
constant level till MJD 53161 and then started increasing steadily till around 
MJD 53165 when the state transition to the thermally dominant state occurred. 
After the transition the flux (and the associated mass accretion rate) started 
increasing rapidly, reaching a peak around MJD 53167.7, and then falling 
sharply again. We do not see any evolution of colors during the LIS, however 
the colors became much softer after the transition, most likely due to the 
formation of an accretion disk. The rms variability also drops significantly 
after the transition. Detailed study is in progress.

Analogous events have been observed in another NS system 4U 1608-52 and few 
black hole systems XTE J1550-564, GRO J0422+32, GRO J1719-24, XTE J1118+480 
as well (see \cite{w2002,ss2005,b2001} and references 
therein) where sometimes the system fails to come out of the hard state during 
the entire outburst. During the 2004 outburst, Aql X-1 entered the regular 
outburst state {\em after} the LIS. The reverse case, when a regular FRED 
outburst occurred {\em before} the LIS, was also seen in the same source during
1998, as shown in Fig.~\ref{lc}(a).

{\bf Conclusions}
\begin{itemize}
\item
The morphology of the optical/X-ray lightcurve is highly non-FRED during  
LIS, for both neutron stars as well as black holes.
\item
The spectral and temporal characteristics of the LIS is similar to that of the 
canonical powerlaw dominated or low/hard state.
\item
The transition from the LIS to an X-ray bright, thermal dominated state is rapid ($<1$ day).
\item
The hard 15-250 keV X-ray flux in this case attained a maximum almost 
simultaneously with the maximum of the optical flux and decayed rapidly as the 
source entered the thermal dominant state.
\end{itemize}

\end{document}